# Deep room impulse response completion

Jackie Lin[1*], Georg Götz[1] and Sebastian J. Schlecht[1,2]

## Abstract

Rendering immersive spatial audio in virtual reality (VR) and video games demands a fast and accurate generation of room impulse responses (RIRs) to recreate auditory environments plausibly. However, the conventional methods for simulating or measuring long RIRs are either computationally intensive or challenged by low signal-to-noise ratios. This study is propelled by the insight that direct sound and early reflections encapsulate sufficient information about room geometry and absorption characteristics. Building upon this premise, we propose a novel task termed "RIR completion," aimed at synthesizing the late reverberation given only the early portion (50 ms) of the response. To this end, we introduce **DECOR**, **D**eep **E**xponential **C**ompletion **O**f **R**oom impulse responses, a deep neural network structured as an encoder-decoder designed to predict multi-exponential decay envelopes of filtered noise sequences. The proposed method is compared against a much larger adapted state-of-the-art network, and comparable performance shows promising results supporting the feasibility of the RIR completion task. The RIR completion can be widely adapted to enhance RIR generation tasks where fast late reverberation approximation is required.

## 1 Introduction

Generating room impulse responses (RIRs) is a well-studied topic with many applications and proposed solutions. A recent application area is virtual acoustics rendering for computer games and augmented and virtual reality (AR/VR), where dynamic sound scenes require realistic and real-time RIRs. Generating RIRs accurately and in real time remains an open task. This paper proposes a new task, *RIR completion*, for fast RIR generation and presents a lightweight deep learning approach, DECOR, that solves this RIR completion task. An evaluation of DECOR shows better performance than an adapted state-of-the-art deep learning model, at a fraction of the size.

*Correspondence:
Jackie Lin
jackiel4@illinois.edu
[1] Acoustics Lab, Department of Information and Communications Engineering, Aalto University, Espoo, Finland
[2] Multimedia Communications and Signal Processing, Friedrich-Alexander-Universität Erlangen-Nürnberg (FAU), Erlangen, Germany

### 1.1 Background

Despite extensive work in RIR generation, challenges remain in broadband accuracy, computational complexity, and real-time synthesis. Room acoustics modeling, such as wave-based and geometrical acoustics, aims to simulate acoustic waves accurately, given a 3D representation of the room with acoustic material assigned to its surfaces. Geometrical acoustics (GA) [1–8], which model sound propagation as a ray, accurately simulate the behavior of high frequencies, but fail to capture wave phenomena such as diffraction especially at low frequencies. Wave-based methods solve the wave equation numerically with methods such as the finite difference time-domain (FDTD) method [9–11], finite element method (FEM) [12], boundary element method (BEM) [13], and spectral element method [14]. Wave-based methods are computationally expensive because complexity exponentially increases with respect to frequency, and quantization and boundary errors cause inaccuracies. The computational complexity of both methods increases considerably with respect to the length of the simulated RIR signal.

Generating broadband RIRs with one single approach is computationally expensive. Therefore, hybrid room acoustics modeling methods for combining the early and the late reverberation, or the high-frequency and low-frequency content from different techniques have been proposed over the past few decades [15–17].

### 1.2 RIR completion

We present the task of RIR completion, where given only the early part of the time-domain RIR (head), the objective is to predict the rest of the RIR sequence (tail), as depicted in Fig. 1. Inspired by hybrid models, the motivation is to develop a lightweight yet dynamic RIR synthesis approach that leverages cheap but accurate computational methods to generate the early portion (i.e., the image source method [2, 3]), and then uses a separate fast RIR completion procedure that generates the late reverb given the aforementioned early portion.

Our primary assumption is that the direct sound and the early reflections in the RIR head contain enough information about the room geometry and acoustic material properties to predict the late reverberation. In the image source method [2, 3], the RIR is synthesized by summing each reflected wavefront that arrives at the receiver with the appropriate distance delay and attenuation. This means that within a short time after excitation, many, if not all, of the room surfaces have reflected energy to the receiver. For example, in a medium-sized rectangular room with dimensions $5\,\mathrm{m} \times 5\,\mathrm{m} \times 3\,\mathrm{m}$, all reflections of second-order and lower, and up to some fourth-order arrive at the receiver within 50 ms. Furthermore, the peaks and times of arrival of early reflections in an RIR are highly consistent across repeated simulations or measurements and can be retrieved more reliably due to a higher signal-to-noise ratio, which supports using the RIR head as reliable information about the room.

To our knowledge, the task of RIR completion, i.e., inferring the RIR tail from just the RIR head, has not been explored. A highly similar task to RIR completion, echo-aware RIR generation, has been concurrently proposed by [19]. We list our contributions as follows:

(I) We propose a lightweight neural network for RIR completion that can process a diverse range of RIR heads from arbitrary rooms, and thus is able to generate late reverberation for dynamic scenes.
(II) We evaluate our proposed method against a state-of-the-art RIR generation approach and show our method achieves a better performance with a much smaller network.

The following section provides an overview of related work in RIR generation using deep learning and examples of similar inverse problems that support the feasibility of our task.

### 1.3 Related work

More recently, the application of deep learning to RIR generation in room acoustics modeling and blind estimation has yielded promising results. To that point, the work of [20] demonstrates that variational autoencoders, and more broadly, deep learning approaches, are well suited for sample-by-sample RIR generation given any informative input (reverberant recordings, geometry, etc.).

Specifically, deep learning approaches for room acoustics modeling have been proposed: Ratnarajah et al. [21] proposed a graph convolution neural network that synthesizes an RIR from the graph representation of an indoor 3D scene. Physics-informed neural networks (PINNs), neural networks constrained by the wave equation, have been proposed for sound field reconstruction and RIR generation [22–24] as an alternative to both traditional wave-based methods and data-driven deep learning methods. Neural representational methods (NeRFs) that encode a room to a continuously queryable representation have been proposed by Luo et al. [25] and

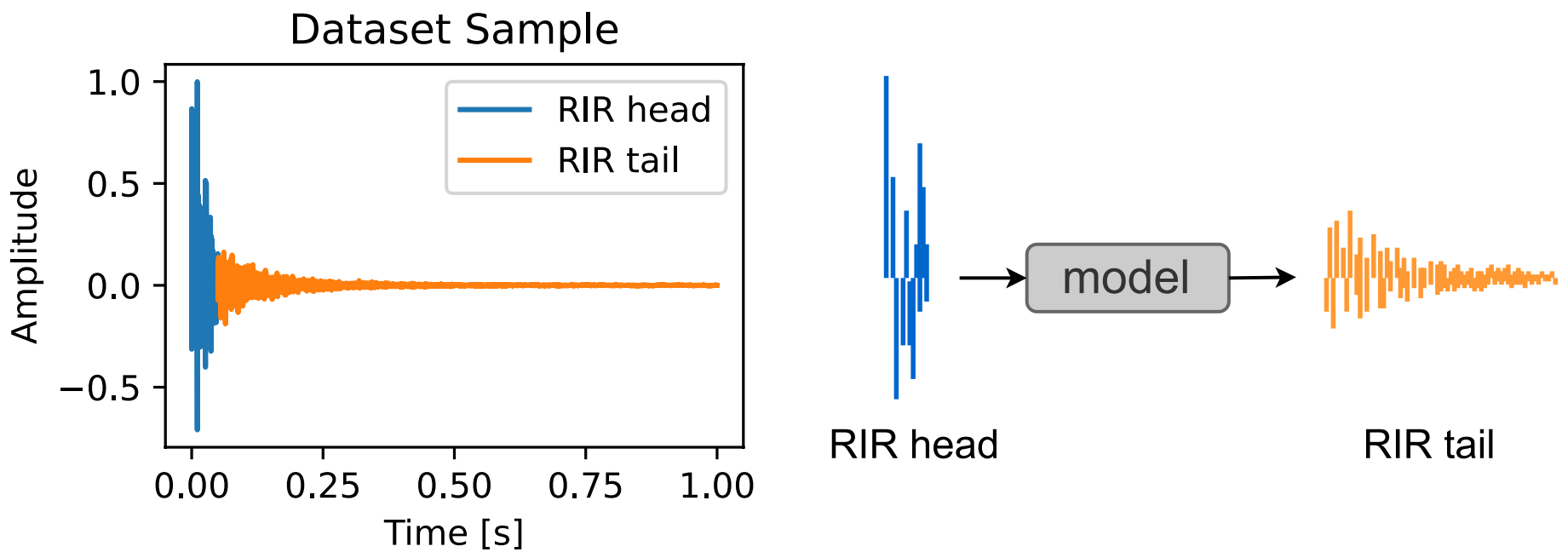

**Fig. 1** Left: example RIR from the Motus [18] dataset, divided at 50 ms into its head and tail components. Right: RIR completion task—take the RIR head and predict the tail

Richard et al. [26] to predict the RIR given the coordinates of the source and receiver.

These examples, however, have limited scope: for example, [21] requires pre-converting the 3D room mesh to its graph representation, taking several seconds, before the graph is input into their neural network. And NeRFs [25, 26] can only output RIRs belonging to the single scene the NeRF model was trained on—each new enclosed space must be represented by a NeRF uniquely trained on RIRs from that space, though work by Majumder et al. [27] and Su et al. [28] attempt to address this shortcoming by inputting additional images and geometry representations of unseen scenes. These methods are not conducive to dynamic AR/VR scenes where room geometry and obstacles may move.

In blind estimation, room parameters or the full RIR signal are inferred from non-RIR input such as reverberant speech recording [29], images [30, 31], or videos of the room [32]. For example, Koo et al. [33] proposed a U-Net model to predict a sample-by-sample RIR given a reverberant singing recording. Similarly, Steinmetz et al. [34] proposed the Filtered Noise Shaping (FiNS) network that is a 1D-convolution encoder-decoder network that takes reverberant speech and predicts a sample-by-sample early part of the RIR (50 ms) and time domain envelopes that shape filtered noise for the late reverberation. We adapt FiNS as a baseline later in the evaluation section.

Lastly, work in geometry prediction using room impulse responses has been explored before, which further supports the tractability of our proposed RIR completion task. Moore [35], Markovic [36], and Kuster [37] used analytical methods to estimate room geometry or volume from a single-channel RIR. Later on, Yu and Kleijn [38] proposed a CNN to estimate the geometry of a room and reflection coefficients from a single RIR. These inverse methods indicate that the RIR contains retrievable information about its corresponding room and scene and thus motivate using the RIR head to predict the RIR tail.

The paper is organized as follows: Section 2 describes our proposed neural network and experimental setup in detail. Section 3 presents the evaluation of our proposed method and compares it to a state-of-the-art RIR generation baseline adapted to the RIR completion task. Section 4 discusses our method's performance on this new task. Section 5 concludes the paper.

## 2 Methods

In this section, we present our neural network **DECOR**, **D**eep **E**xponential **C**ompletion **O**f **R**oom impulse responses, that takes the RIR time domain head and predicts the RIR tail, i.e.,

$$\Phi : \boldsymbol{h}_{\text{head}\,[0\,\text{s},\,50\,\text{ms}]} \rightarrow \hat{\boldsymbol{h}}_{\text{tail}\,[50\,\text{ms},\,1\,\text{s}]}\,. \quad (1)$$

First, we present the encoder-decoder structure of DECOR, shown in Fig. 2, with a detailed explanation of the acoustics-informed decoder. Then, we discuss the loss function, datasets, and experimental setup details that were used in the training of the model.

### 2.1 Encoder

We modify the encoder structure from the FiNS model [34] which originally takes a few seconds of reverberant speech, for the short RIR head input. The DECOR time-domain encoder takes the first 50 ms of the RIR sampled at 48 kHz as the input $\boldsymbol{h}_{\text{head}} \in \mathbb{R}^{2400}$ and performs a series of strided 1D convolutions and skip connections via the encoding block described in [34]. Nine encoding blocks progressively downsample $\boldsymbol{h}_{\text{head}}$. The output is passed through an adaptive 1D pooling layer and then through a single linear layer to obtain the latent vector $\boldsymbol{z}$ with desired embedding length $k = 128$.

### 2.2 Decoder

We designed the decoder with strong room acoustics inductive bias to minimize the model size and amount of training data while maximizing expressive power. The decoder of DECOR is based on the exponentially decaying white noise model of reverberation, described by

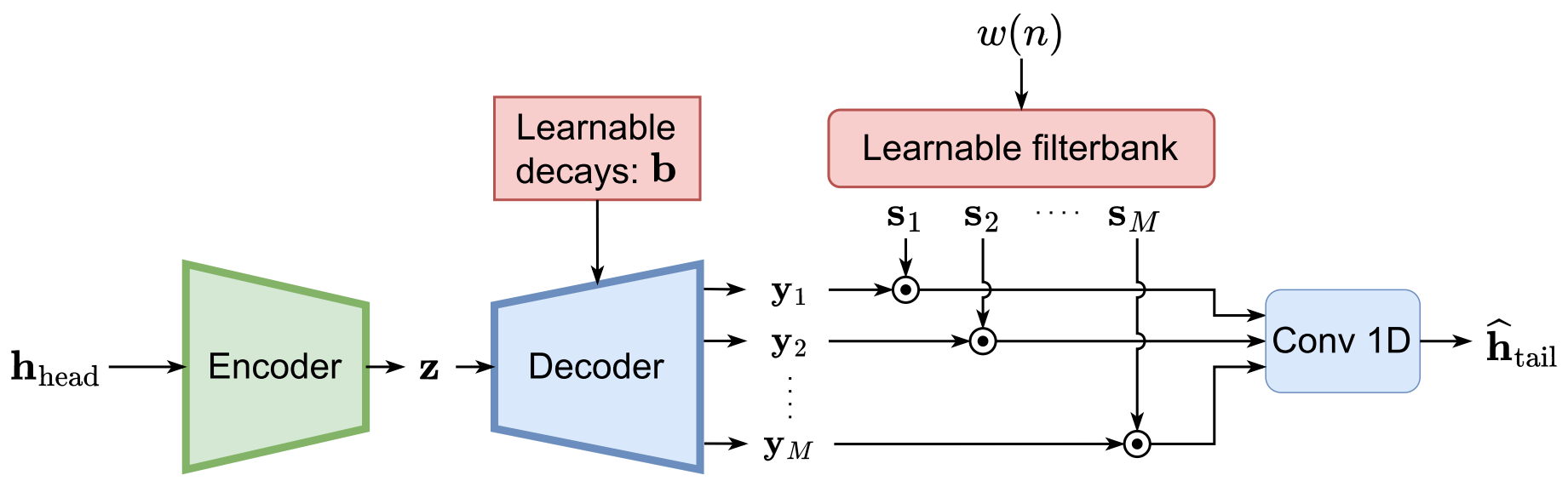

**Fig. 2 DECOR** overview. The RIR head $\boldsymbol{h}_{\text{head}}$ is passed through an encoder-decoder architecture. Within the decoder, we predict multi-exponential decay envelopes $\boldsymbol{y}_i$, which are used to shape filtered noise sequences $\boldsymbol{s}_i$. The shaped noise sequences are combined to form the RIR tail $\hat{\boldsymbol{h}}_{\text{tail}}$

Moorer [39]. Due to the stochastic nature of late reverberation, the room impulse response $h(n)$ can be modeled as stochastic white noise $w(n)$ enveloped by a sum of $N$ exponential decays

$$h(n) = w(n) \sum_{j=1}^{N} a_j e^{-b_j n}, \quad (2)$$

where $a_j$ and $b_j$ are the initial amplitude of the decay and decay rate, respectively.

This impulse response representation can be further broken down into frequency bands $i = 1, ..., M$ to capture frequency-dependent decay,

$$h(n) = \sum_{i=1}^{M} h_i(n) = \sum_{i=1}^{M} s_i(n) \sum_{j=1}^{N} a_{ij} e^{-b_j n}, \quad (3)$$

where $s_i(n)$ is band-limited noise corresponding to the different frequency bands.

For a discrete-time sequence of length $T$, we omit $n$ and simplify the notation to

$$\boldsymbol{h} = \mathbf{1}_M^{\mathsf{T}} \boldsymbol{S} \odot \boldsymbol{Y}, \quad (4)$$

where $\boldsymbol{Y} \in \mathbb{R}^{M \times T}$ are the time-domain envelopes for the filtered white noise sequences $\boldsymbol{S} \in \mathbb{R}^{M \times T}$, $\odot$ denotes element-wise multiplication, and a column vector of ones $\mathbf{1}_M^{\mathsf{T}}$ sums the column elements. The time-domain envelopes $\boldsymbol{Y}$ can be expressed in vector notation as linear combinations of exponential decay envelopes

$$\boldsymbol{Y} = \boldsymbol{A}\boldsymbol{E}, \quad (5)$$

where $\boldsymbol{A} \in \mathbb{R}^{M \times N}$ are the initial amplitudes of the exponential decay envelopes $\boldsymbol{E}$ for $N$ decay rates $\boldsymbol{b} \in \mathbb{R}^N$ over time sequence $\boldsymbol{n} \in \mathbb{R}^T$, i.e.,

$$\boldsymbol{E} := e^{-\boldsymbol{b}\boldsymbol{n}^{\mathsf{T}}} \in \mathbb{R}^{N \times T}. \quad (6)$$

The proposed decoder is depicted in Fig. 3. It constructs the time-domain envelopes $\boldsymbol{Y}$ by predicting the decay envelope amplitude values $\boldsymbol{A}$ and multiplying them with the exponential decay envelopes $\boldsymbol{E}$.

A multihead MLP with seven hidden layers takes the latent vector $\boldsymbol{z}$ and outputs two matrices $\boldsymbol{A}'$ and $\boldsymbol{C}'$. The element-wise product of $\boldsymbol{A}'$ and the mask $\boldsymbol{C} = \sigma(\boldsymbol{C}')$ gives the decay envelope amplitude matrix

$$\boldsymbol{A} = \boldsymbol{A}' \odot \sigma(\boldsymbol{C}'), \quad (7)$$

where $\sigma(\cdot)$ is the sigmoid function and $\boldsymbol{C} = \sigma(\boldsymbol{C}') = \{c_{ij} \mid 0 \leq c_{ij} \leq 1\}$.

We found that learning a linear representation $\boldsymbol{A}'$ and a sigmoid mask achieved better results than learning $\boldsymbol{A}$ directly. The separate prediction of $\boldsymbol{C}'$ and application of the sigmoid function yields a mask that enforces close to zero amplitudes for non-active decays. This strategy helped to enforce sparsity in $\boldsymbol{A}$, see Fig. 7.

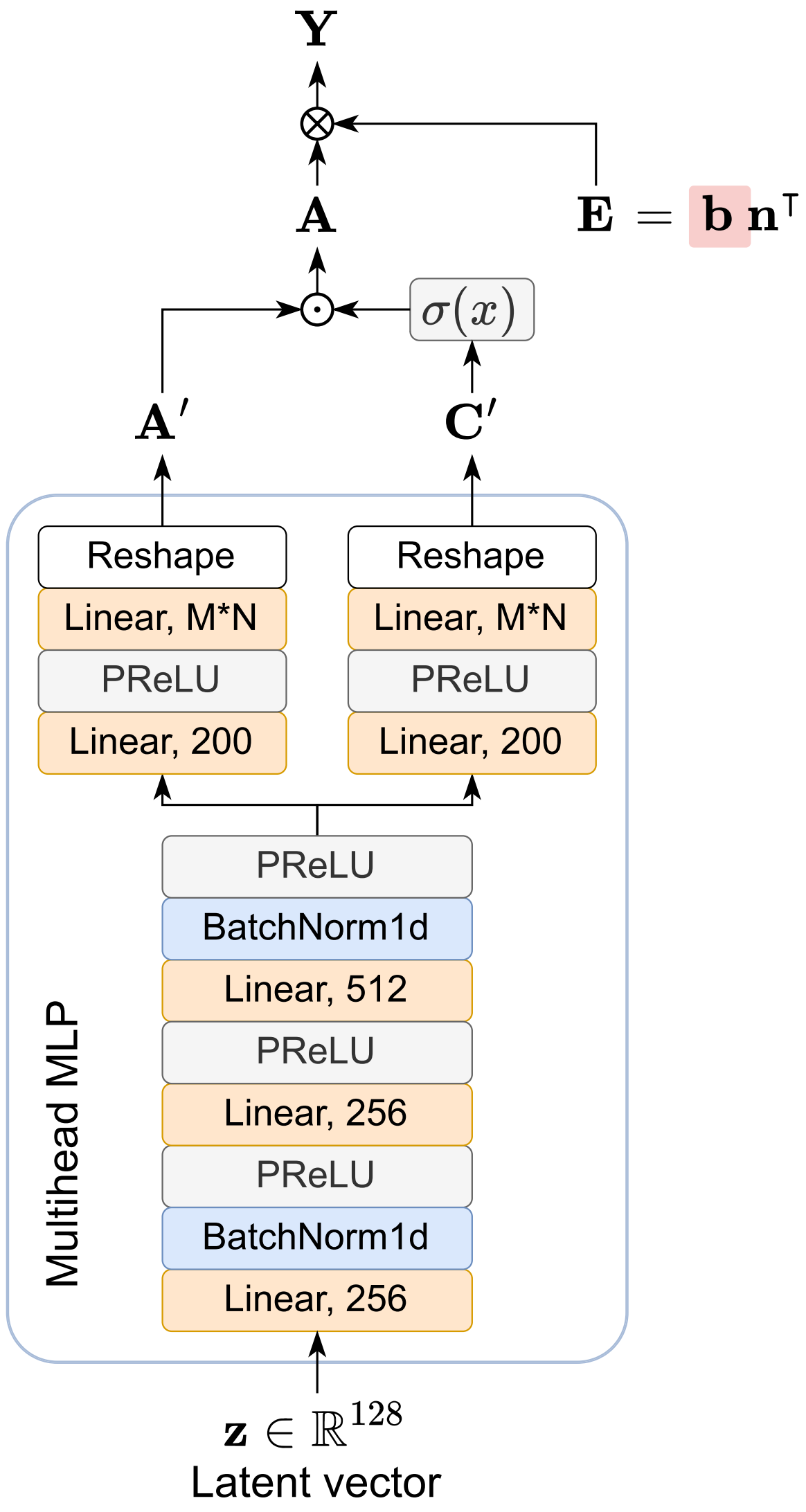

**Fig. 3** Decoder structure. The latent vector $\boldsymbol{z}$ is fed through a multihead MLP, producing $\boldsymbol{A}'$ and mask $\sigma(\boldsymbol{C}')$. The exponential decay envelopes $\boldsymbol{E}$ are constructed separately, using the learnable decay rates $\boldsymbol{b}$. The amplitude matrix is calculated as $\boldsymbol{A} = \boldsymbol{A}' \odot \sigma(\boldsymbol{C}')$ and multiplied with $\boldsymbol{E}$ to output the time-domain decay envelopes $\boldsymbol{Y}$

The exponential decay envelopes $\boldsymbol{E}$ is then constructed from $\boldsymbol{b}$ and $\boldsymbol{n}$ using Eq. (6). The learnable parameter $\boldsymbol{b} = \{b_j \mid b_j = \ln(10^{-3})/T_j\} \in \mathbb{R}^N$ is initialized before training and fixed during inference, with $T_j$ being the T60 decay time of the $j$th slope. In our final model, we initialized $N = 20$ decay times, logarithmically sampled from the range 0.05 to 3.0 s. The envelope matrix $\boldsymbol{E}$ is calculated for a 950 ms time sequence corresponding to the RIR tail, i.e., $\boldsymbol{n} = [0.05, 0.05 + 1/f_s, ..., 1.0] \in \mathbb{R}^T$, with sequence length $T = 45600$ at sampling rate 48 kHz.

Taking a similar filtered noise approach as in FiNS [34], we construct a filterbank of $M$ learnable FIR filters that processes a Gaussian white noise signal $\boldsymbol{w} \in \mathbb{R}^T$ into $M$ filtered noise signals $\boldsymbol{S} = [\boldsymbol{s}_1, \boldsymbol{s}_2, ...\boldsymbol{s}_M]$, see Fig. 2. We initialize the learnable filterbank with $M = 10$ FIR octave band filters of order $P = 1023$. Finally, the RIR signal is constructed from the element-wise product of the time-domain envelopes and the filtered noise signals

$$\boldsymbol{H} = \boldsymbol{S} \odot \boldsymbol{Y}, \tag{8}$$

where the rows of $\boldsymbol{H} \in \mathbb{R}^{M \times T}$ are filtered noise signals weighted with the learned envelopes. A last 1D convolution layer linearly combines the rows of $\boldsymbol{H}$ to return the full-band predicted RIR tail

$$\hat{\boldsymbol{h}}_{\text{tail}} = \boldsymbol{w}_{\text{conv_1D}}^{\mathsf{T}} \boldsymbol{H}. \tag{9}$$

## 2.3 Experimental setup

In the following, we present the loss function, dataset, and training parameters.

### 2.3.1 Loss function

A multiresolution short-time Fourier transform (MSTFT) loss function [34, 40] was used to train and evaluate the model. The MSTFT loss function $\mathcal{L}_{\text{MSTFT}}(\hat{h}, h)$ between the predicted RIR $\hat{h}(n)$ and the true RIR $h(n)$ is given as the sum of $R$ STFT losses with different STFT resolutions, i.e.,

$$\mathcal{L}_{\text{MSTFT}}(\hat{h}, h) = \sum_{r=1}^{R} \left[ \mathcal{L}_{\text{sc},r}(\hat{h}, h) + \mathcal{L}_{\text{sm},r}(\hat{h}, h) \right]. \tag{10}$$

The STFT loss is the sum of the spectral convergence loss

$$\mathcal{L}_{\text{sc},r}(\hat{h}, h) = \frac{\left\| |\text{STFT}_r(h)| - |\text{STFT}_r(\hat{h})| \right\|_{\text{F}}}{\left\| |\text{STFT}_r(h)| \right\|_{\text{F}}} \tag{11}$$

and the spectral log-magnitude loss

$$\mathcal{L}_{\text{sm},r}(\hat{h}, h) = \frac{1}{N} \| \log\left(|\text{STFT}_r(h)|\right) - \log\left(|\text{STFT}_r(\hat{h})|\right) \|_1, \tag{12}$$

where $\|\cdot\|_{\text{F}}$ is the Frobenius norm, and $\|\cdot\|_1$ is the $L_1$ norm.

During our training, we used $R = 4$ resolutions, with window sizes [64, 512, 2048, 8192], hop sizes [32, 256, 1024, 4096], and Hann windowing.

### 2.3.2 Dataset

We trained DECOR on 36,555 RIRs over 256 and 5094 unique measured and simulated rooms respectively, combined from five public datasets: the Arni dataset [41], R3VIVAL dataset [42], Motus dataset [18], the MIT Acoustical Reverberation Scene Statistics Survey [43], and the GWA dataset [44]. Additionally, the BUT Reverb Database [45] is not seen during training, and it is used to test the generalization ability of DECOR. Details of the datasets are shown in Table 1.

All RIRs from Motus, MIT Survey, and R3VIVAL are used. From Arni, a uniformly random sampled subset of variable acoustic panel configurations is used to maintain proportional representation in the training data. RIRs from the same configuration (Arni, Motus, and R3VIVAL) were kept solely within their designated train, validate, or test dataset, so all validation and test RIRs are from unseen configurations. For the GWA dataset, RIRs with very low energy are discarded, and the GWA dataset is only used for training.

The data is preprocessed to ensure consistency across datasets. In the case of an ambisonics audio file, the omnidirectional channel is chosen. Otherwise, a random

**Table 1** Datasets used in the training, validation, testing of the model, and generalization to unseen datasets

| Use | Dataset name | Type | # RIRs | # rooms | # unique configs | Description |
|---|---|---|---|---|---|---|
| Training + validation + testing | Arni [41] | Measured | 2240 | 1 | 2240 | 48 kHz. Mono-channel. Variable acoustics room |
| | Motus [18] | Measured | 3320 | 1 | 830 | 48 kHz. 4th-order ambisonics. Variable furniture and wall covering materials. 360 ° photo per room configuration |
| | MIT Survey [43] | Measured | 271 | 271 | 271 | 32 kHz. Mono-channel |
| | R3VIVAL [42] | Measured | 272 | 1 | 8 | 192 kHz. SRIRs (using SDM). Variable acoustics room. 360 ° video |
| | GWA dataset [44] | Simulated | 31,429 | 5661 | 5661 | 48 kHz. Mono-channel. Geometrical acoustics and wave-based hybrid simulation |
| Total | | | 36,555 | 5935 | 9010 | |
| Generalization | BUT ReverbDB [45] | Measured | 2325 | 9 | 9 | 16 kHz. Mono-channel |

The number of unique configurations refers to the number of unique absorption scenarios; e.g., the Arni dataset has 5342 unique configurations of variable acoustics wall panels channel is chosen. All files are resampled to 48 kHz. We normalize each RIR to absolute amplitude 1.0 and remove any initial delay. Finally, we separate the RIR into the 50 ms head $[0, 50\,\text{ms}]$ and the 950 ms tail $[50\,\text{ms}, 1\,\text{s}]$ portion, corresponding to the neural network input and target. The model is trained to be robust to the sampling rate. This is achieved by applying a lowpass filter with a randomly chosen cutoff frequency $f_{\text{cutoff}} = \{8\text{k}, 12\text{k}, 16\text{k}, 22.05\text{k}, 24\text{k}\}$ to each RIR in a minibatch batch during training. The train-valid-test split is [36,555, 439, 537] RIRs.

#### 2.3.3 Training parameters

During training, we used the Ranger21 [46] optimizer (based on the AdamW optimizer) with an initial learning rate of $1 \times 10^{-4}$. The model was trained for 1000 epochs with a batch size of 128 on an A100 GPU, which took 23 h.

## 3 Results

We present the results of the proposed method against a baseline method using various evaluation metrics.

### 3.1 Baseline

We construct a deep learning baseline by modifying FiNS [34] for our RIR completion task. FiNS was originally used for the blind estimation task. Therefore, we adapt the FiNS encoder as described in Section 2.1, while preserving the FiNS decoder, which uses a convolution upsampling approach. We note that our model is more than 30 times smaller than the FiNS baseline, at 37 MB vs. 1.3 GB, respectively.

Additionally, we construct a naive signal processing baseline to give context to the error magnitudes of the proposed and FiNS baseline methods. This naive baseline uses a stochastic RIR, modeled as white noise shaped with an exponential decay envelope, with T60 and energy matching the mean T60 (0.65 s) and energy of the training data. The error of naive baseline, consisting of this single representative RIR of the training dataset, is calculated relative to the ground truth RIRs in the test dataset.

### 3.2 Evaluation metrics

We use four objective metrics to evaluate the performance of the models: MSTFT error, energy decay function (EDF) error, reverberation time (T60) error, and direct-to-reverberant ratio (DRR) error. These metrics give a rough indication of the ability to match the target acoustic room characteristics.

MSTFT error [see (10)] evaluates the similarity of two RIR spectra across multiple STFT resolutions [34, 40]. Besides using it as the loss function during the training phase, it continues to be informative as an evaluation metric as it considers spectral and temporal differences in the STFT domain.

EDF error is the error between the predicted and the true EDF [47]. The mean absolute error (MAE) (13) and the root mean squared error (RMSE) (14) are reported

$$\text{EDF}_{\text{MAE}} = \frac{1}{T}\sum_{n=1}^{T} \hat{d}^{(\text{dB})}(n) - d^{(\text{dB})}(n) \tag{13}$$

$$\text{EDF}_{\text{RMSE}} = \sqrt{\frac{1}{T}\sum_{n=1}^{T} \left[\hat{d}^{(\text{dB})}(n) - d^{(\text{dB})}(n)\right]^2} \tag{14}$$

where the EDFs $\hat{d}^{(\text{dB})}(n)$ and $d^{(\text{dB})}(n)$ are computed using the Schroeder backward integration procedure [48] and represented on a logarithmic scale in dB. EDF error quantifies how much the sound energy decay differs between the true and predicted RIR.

T60 error is a widely used metric to evaluate reverberation generation, as it broadly captures similarity in reverberation time, which is a perceptual cue for room size and acoustic absorption. T60 error is the mean absolute percentage error (MAPE) between the true $T60_c$ and predicted $\widehat{T60}_c$ reverberation time over $n$ octave bands, i.e.,

$$\text{T60}_{\text{MAPE}} = 100\frac{1}{n}\sum_{c=1}^{n} \left| \frac{\widehat{T60}_c - T60_c}{T60_c} \right| . \tag{15}$$

In our evaluation, we use the DecayFitNet proposed by Götz et al. [47] to determine the T60s of the ground truth and predicted RIRs for octave bands with center frequencies $f_c = \{125, 250, 500, 1000, 2000, 4000\}$.

The final metric used is the DRR error, which is the MSE between true and predicted DRR. The DRR is an energy ratio, computed as the energy of direct sound divided by the energy of late reverberation. It captures the perception of source distance and sense of reverberance and is also commonly used as an RIR evaluation metric. DRR is computed from the RIR as

$$\text{DRR} = 10 * \log_{10}\left(\frac{\sum_{n=n_d-n_0}^{n_d+n_0} h^2(n)}{\sum_{n=n_d+n_0}^{\infty} h^2(n)}\right) \tag{16}$$

where $n_d$ is the sample index of the direct sound peak and $n_0$ is the number of samples corresponding to a small temporal window of 1 ms.

### 3.3 Performance evaluation

The DECOR model and the FiNS baseline successfully predict RIR tail from the RIR head, as indicated by the example shown in Fig. 5. The models correctly estimate the temporal and spectral characteristics of the tail, and also the sound decay behavior shows good agreement

**Table 2** Test error

| Model | MSTFT (↓) | EDF (MAE, dB, ↓) | EDF (RMSE, dB, ↓) | T60 (MAPE, %, ↓) | DRR (MSE, dB, ↓) |
|---|---|---|---|---|---|
| Naive | 2.60 | 10.4 | 13.0 | 29.0 | 13.9 |
| FiNS | 1.05 | 5.21 | 7.56 | 20.1 | 1.69 |
| DECOR | **0.97** | **4.04** | **6.19** | **14.6** | **1.15** |

The model was trained on 36.5k RIRs from a combined dataset of Arni [41], R3VIVAL [42], MOTUS [18], the MIT Survey [43], and GWA dataset [44]. MSTFT error, EDF MAE and RMSE, T60 MAPE, and DRR MSE are reported. See Section 3.2 for a formulation of the evaluation metrics. Arrows indicate lower values for the metrics are better

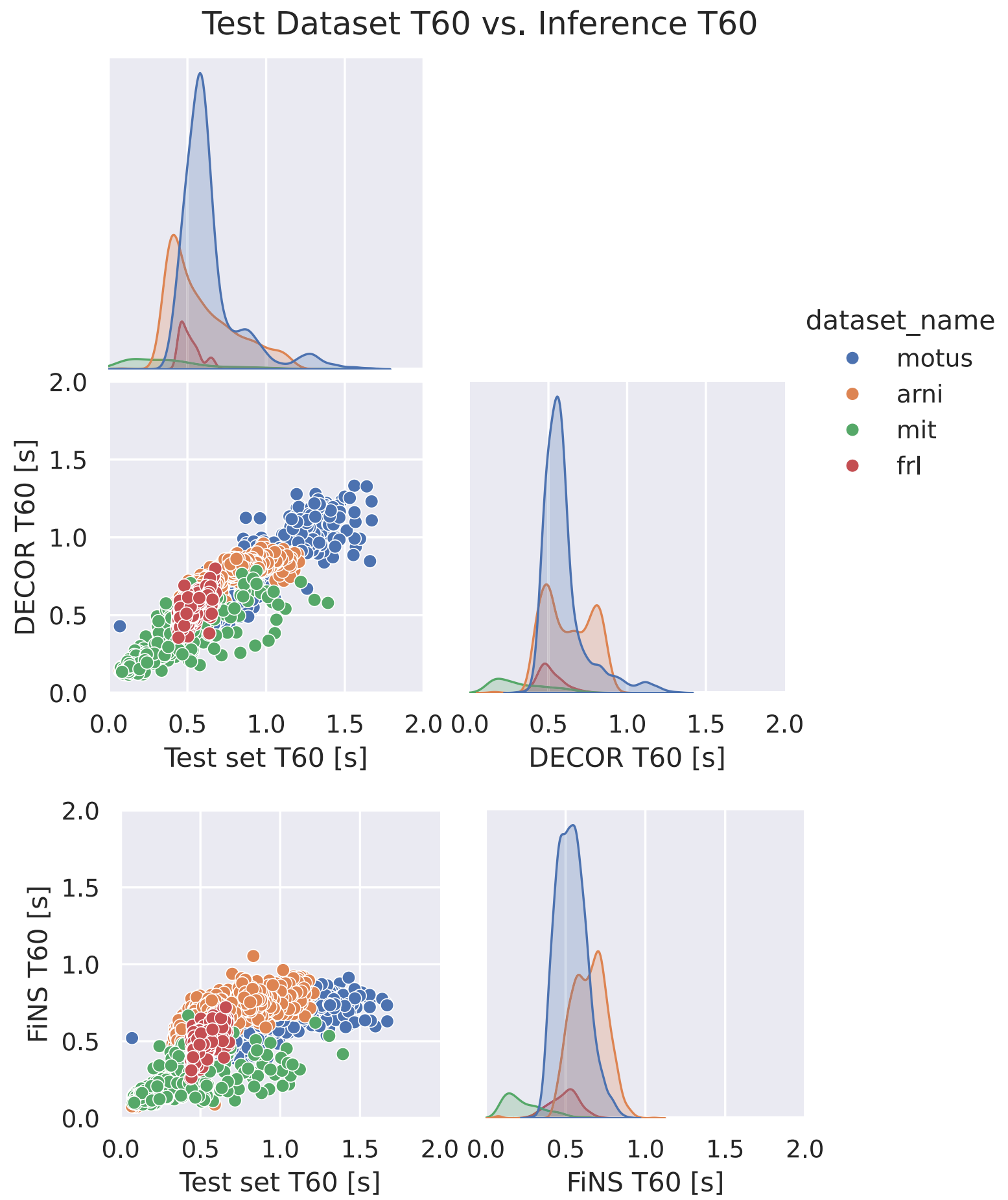

**Fig. 4** Test dataset T60 vs. inference T60. Top row is test dataset T60 distribution. Middle row is DECOR inference. Bottom row is FiNS baseline inference. Perfect T60 match is when all points fall on the identity function $x = y$

with the ground truth. Table 2 summarizes the evaluation metrics on the test dataset, which contains unseen rooms from the training datasets shown in Table 1. DECOR performs better on all metrics compared to the deep learning baseline.

Examining the models further, both DECOR and the baseline slightly underestimate T60, but DECOR has better agreement with the true T60s, as shown in Fig. 4. On average, DECOR inferred RIRs within 14.6% of the ground truth T60, which is slightly above the threshold of T60 just noticeable differences [49].

We conducted an informal perceptual evaluation of our model. Sound examples can be found on the project website[1]. Both models produced RIRs closely matching the timbre of the ground truth room. However, the FiNS baseline model generated unnatural-sounding RIRs. The waveform synthesized by the baseline contains sparse, large amplitude peaks, as illustrated on the bottom left of Fig. 5. These peaks add an unnatural graininess that is not present in the ground truth RIR tail. In contrast, our predicted RIRs achieve much smoother-sounding RIRs. This is because DECOR predicts parameters for the exponential decay model in Eq. (3), thus generating smoothly decaying RIRs with exponential decay curves. When convolved with music or speech, the differences are less apparent.

[1] Website: https://linjac.github.io/rir-completion/

### 3.4 Generalization power

Lastly, we evaluated DECOR on an unseen, measured RIR dataset to investigate its generalization power. We use the BUT ReverbDB [45] dataset, and the corresponding error values are reported in Table 3.

The reported values across all metrics indicate that our model performs worse on an unseen dataset than on the test dataset (Motus, Arni, R3VIVAL, MIT Survey). The baseline model also performs worse, and DECOR outperforms the baseline on EDF MAE, T60, and DRR metrics. Both models perform better than a naive guess. Both DECOR and the baseline significantly underestimate unseen dataset T60s but positively correlates with the true values, shown in Fig. 6. The slight positive relationship indicates that DECOR is able infer non-random and relatively correct T60s.

## 4 Discussion

The results show that DECOR performed successfully on the RIR completion task. It generated a realistic RIR tail with temporal, spectral, and sound energy decay characteristics matching the ground truth. Our model performs

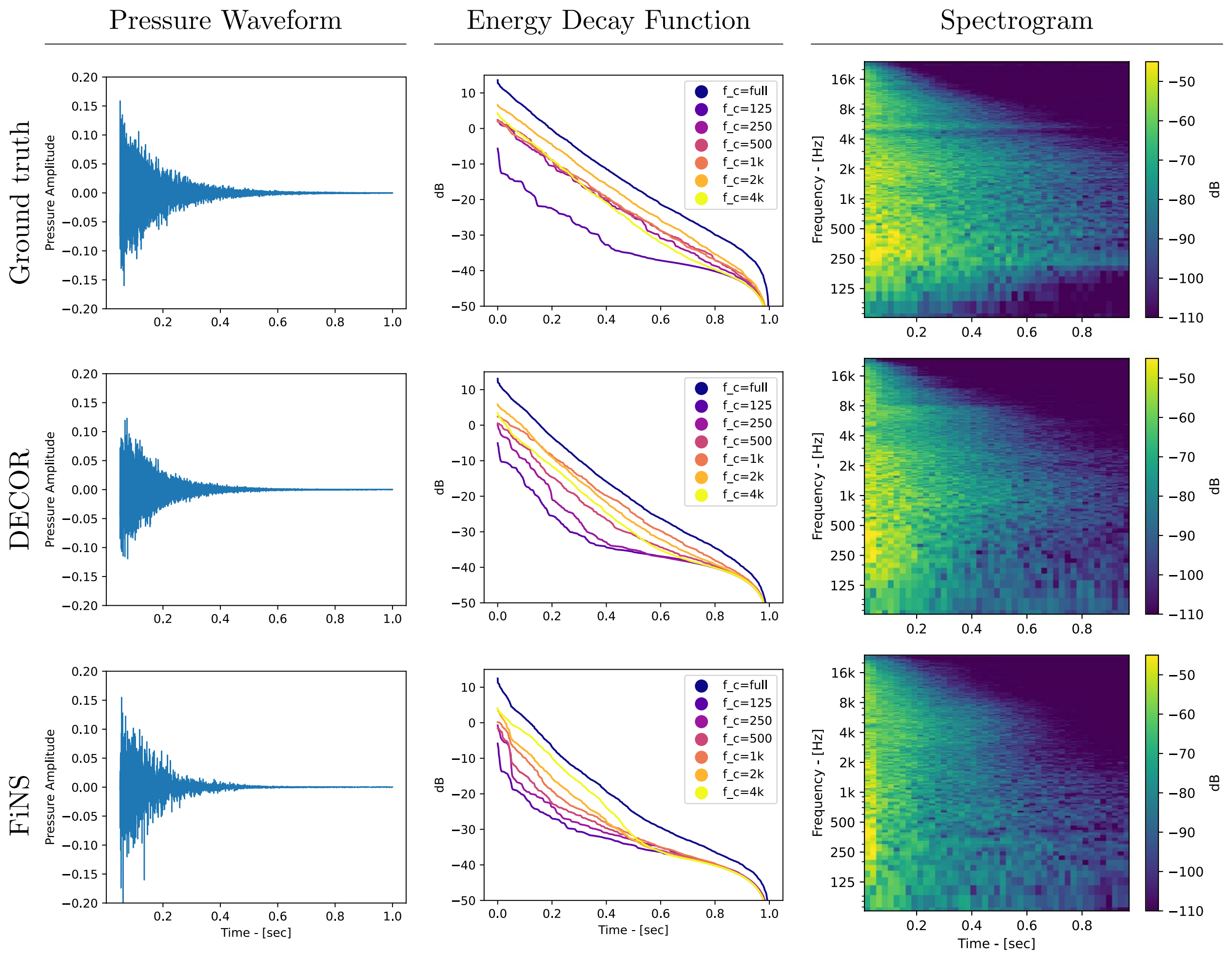

**Fig. 5** Model outputs on a test dataset sample. The top row is the ground truth; the middle row is the DECOR model, and the bottom row is the FiNS baseline. The left column shows the RIR tail waveform. The middle column shows the unnormalized EDF from the full-length broadband signal (darkest) and in octave bands (dark to light with increasing band center frequency). The right column shows the magnitude spectrogram of the full-length RIR

**Table 3** Generalization power: error on an unseen dataset, BUT ReverbDB [45], which contains 1.3k RIRs measured in 9 unique rooms

| Model | MSTFT (↓) | EDF (MAE, dB, ↓) | EDF (RMSE, dB, ↓) | T60 (MAPE, %, ↓) | DRR (MSE, dB, ↓) |
|---|---|---|---|---|---|
| Naive | 3.53 | 12.8 | 17.3 | 41.6 | 14.5 |
| FiNS | **1.32** | 10.9 | **13.9** | 41.9 | 8.68 |
| DECOR | 1.42 | **10.7** | 14.0 | **39.5** | **8.22** |

Arrows indicate lower values for the metrics are better

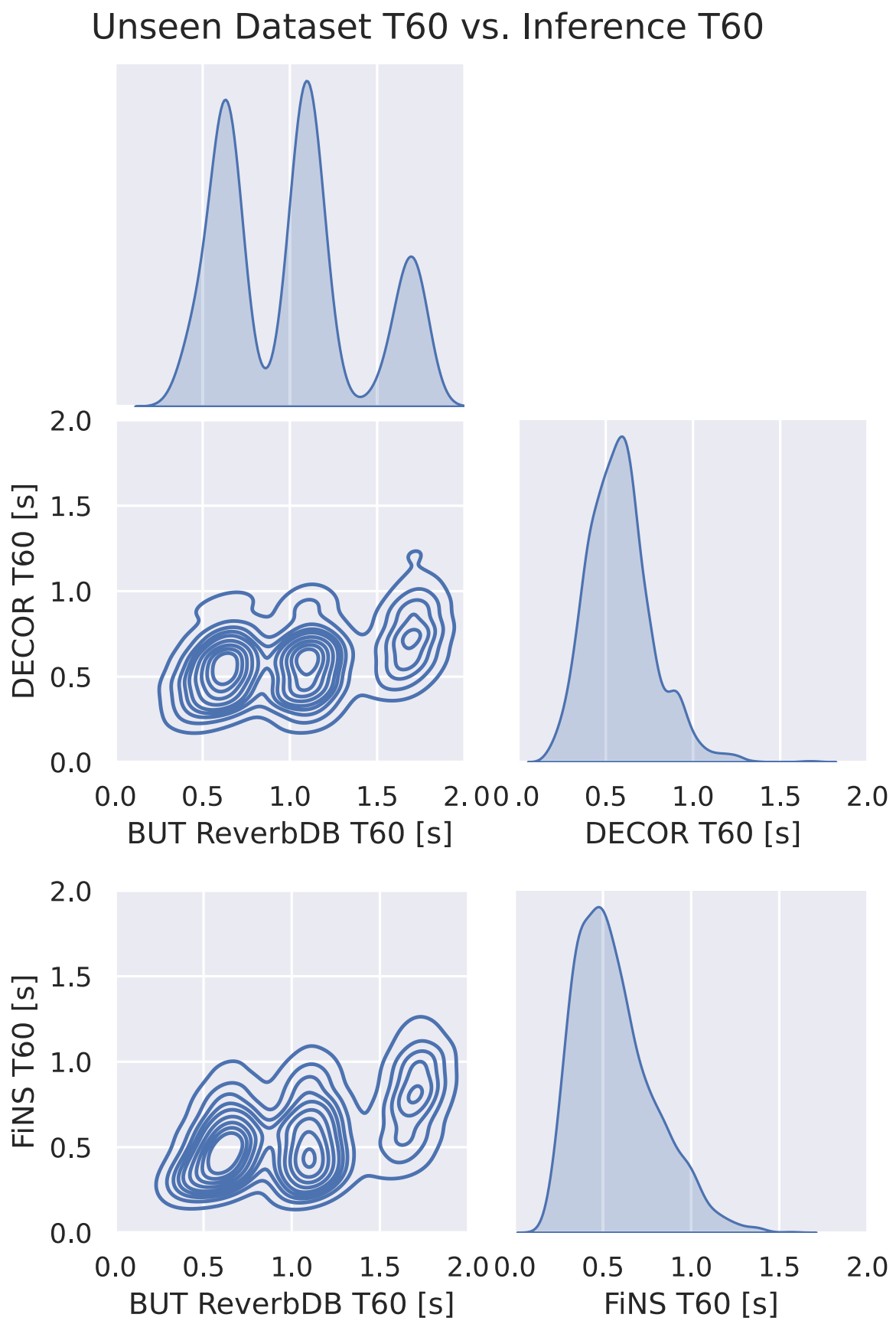

**Fig. 6** Unseen dataset T60 (BUT ReverbDB) vs. inference T60. Contour lines indicate sample density. Top row is unseen dataset T60 distribution. Middle row is DECOR inference. Bottom row is FiNS baseline inference. Perfect T60 match is when contour lines collapse to the identity function $x = y$

better than the baseline and is more than thirty times smaller.

Secondly, the results show that the DECOR model's performance decreases significantly when evaluating it on an unseen dataset. A similar loss of generalization performance was already noted in the original FiNS paper [34]. Our training dataset only consisted of five public datasets; even with few datasets, DECOR is able to learn a proportional relationship between the ground truth versus inferred T60 [see Fig. 6]. Increasing and diversifying the types of rooms and datasets during training will likely improve the generalization ability.

Our model achieves good results in the context of RIR completion. Extrapolating a signal to twenty times its original length can be considered a challenging task. The difficulty of the task helps to contextualize the network's performance, for example, when comparing it with results reported in RIR blind estimation performance of the original FiNS model [34], where on their test set, MSTFT error was 1.18, cf. Table 2]. The informal perceptual evaluation also supports the idea that our synthesized RIRs sound similar to the ground truth regarding timbre, reverberation time, and DRR.

One possible use case for the RIR completion task is real-time RIR generation for augmented and virtual reality (AR/VR), thus motivating this approach with low computational complexity and storage requirements. The DECOR encoder-decoder performs a general regression task on the RIR head to determine the activation of a range of decay rates, see Fig. 7. DECOR's amplitude matrix $\boldsymbol{A}$ representation of the RIR lies in between the multi-slope decay model [47, 50] and the damping density model introduced by Kuttruff [7]; it follows that DECOR is interpretable and highly-expressive with the ability to express the reverberation of complicated scenarios such as coupled rooms with multiple decay slopes.

## 5 Conclusion

In this paper, we propose a deep neural model DECOR that closely approximates an RIR given the short beginning segment. The performance of the DECOR model is better than a baseline method. It is more than thirty times smaller, has fewer noticeable artifacts, and is highly expressive. DECOR was found to perform worse on an unseen dataset, but indicates generalization ability. While the proposed method has room for improvement

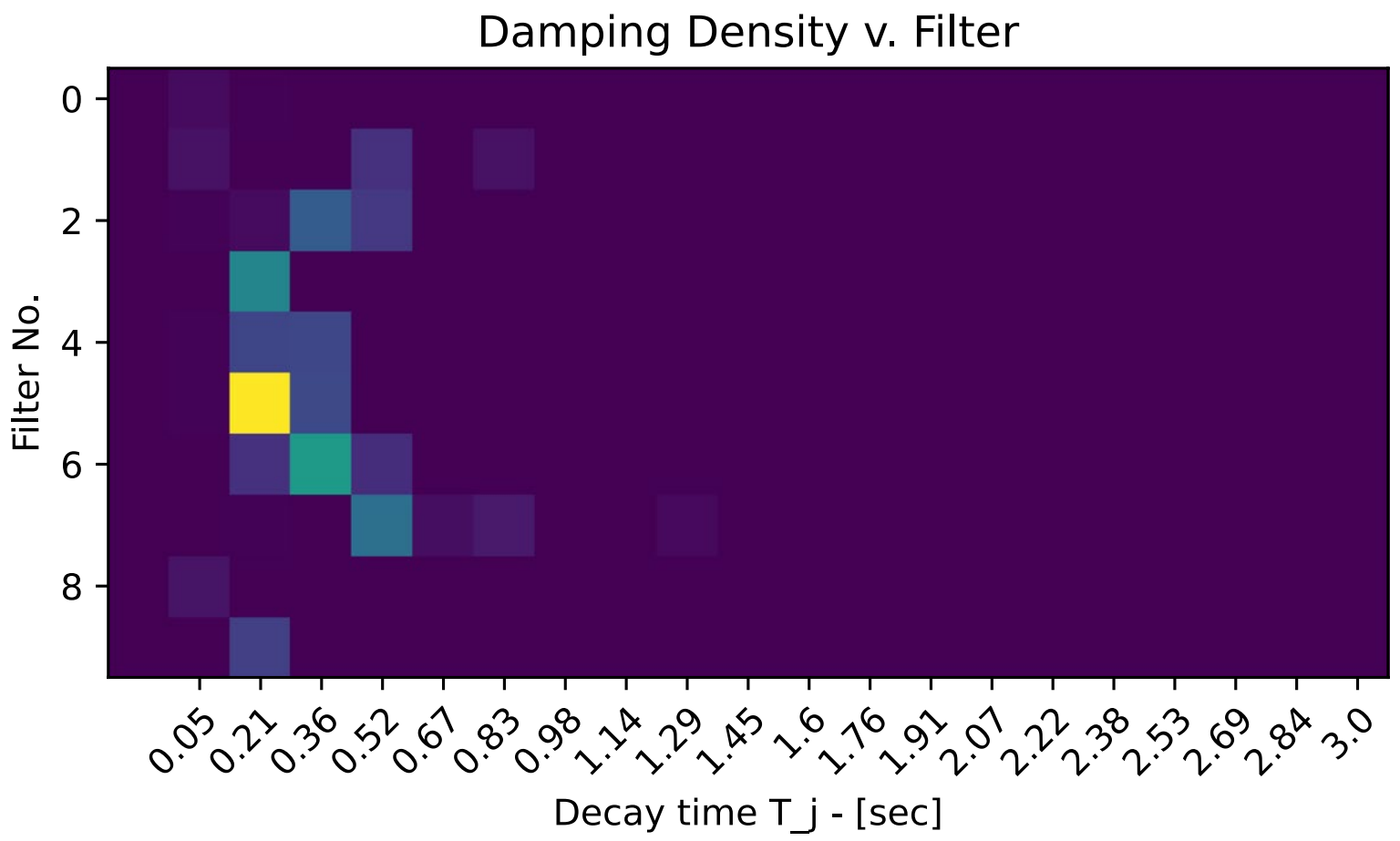

**Fig. 7** Amplitude matrix ***A*** for a predicted RIR. The model predicts one dominant decay time per filter, while slightly activating adjacent decay times, and inactivating all other decay times. Note that the dominant decay is different per filter

and exploration, for example, improving generalization ability and extending capability to variable input length, this work demonstrates that RIR completion is a solvable task. RIR completion offers a promising new direction for applications that require the fast generation of the late part of room impulse responses.

**Acknowledgements**
This work was done at the Acoustics Lab at Aalto University School of Electrical Engineering. We would like to thank Ricardo Falcon Perez and Kyungyun Lee for their invaluable input and guidance on designing and running deep learning experiments. We acknowledge the computational resources provided by the Aalto Science-IT project.

**Authors' contributions**
Jackie Lin conceptualized the publication and implemented the proposed method, ran experiments, produced plots, and wrote the article. Sebastian Schlecht and Georg Götz contributed key ideas, advised the work, and revised the article. All authors reviewed and approved the final manuscript.

**Funding**
Jackie Lin was funded by the Aalto University School of Electrical Engineering, Funding for multidisciplinary MSc theses.

**Data availability**
All data used in the training and evaluation of the deep learning models in this paper are from publicly available datasets. DOIs for all datasets are included in the reference list.

## Declarations

**Competing interests**
The authors declare that they have no competing interests.

**References**

1. L. Savioja, U.P. Svensson, Overview of geometrical room acoustic modeling techniques. J. Acoust. Soc. Am. **138**(2), 708–730 (2015). https://doi.org/10.1121/1.4926438
2. J.B. Allen, D.A. Berkley, Image method for efficiently simulating small-room acoustics. J. Acoust. Soc. Am. **65**(4), 943–950 (1979). https://doi.org/10.1121/1.382599
3. J. Borish, Extension of the image model to arbitrary polyhedra. J. Acoust. Soc. Am. **75**(6), 1827–1836 (1984). https://doi.org/10.1121/1.390983
4. A. Krokstad, S. Strøm, S. Sørsdal, Calculating the acoustical room response by the use of a ray tracing technique. J. Sound Vib. **8**(1), 118–125 (1968). https://doi.org/10.1016/0022-460x(68)90198-3
5. A. Kulowski, Algorithmic representation of the ray tracing technique. Appl. Acoust. **18**(6), 449–469 (1985). https://doi.org/10.1016/0003-682X(85)90024-6
6. M. Vorländer, Simulation of the transient and steady-state sound propagation in rooms using a new combined ray-tracing/image-source algorithm. J. Acoust. Soc. Am. **86**(1), 172–178 (1989). https://doi.org/10.1121/1.398336
7. H. Kuttruff, *Room Acoustics*, 5th edn. (CRC Press, New York, 2009)
8. T. Lewers, A combined beam tracing and radiatn exchange computer model of room acoustics. Appl. Acoust. **38**(2–4), 161–178 (1993). https://doi.org/10.1016/0003-682X(93)90049-C
9. L. Savioja, T. Rinne, T. Takala, , in *Proc. Int. Computer Music Conf.* Simulation of room acoustics with a 3-D finite difference mesh (Aarhus, 1994), International Computer Music Association ICMA, pp. 463–466
10. D. Botteldooren, Finite-difference time-domain simulation of low-frequency room acoustic problems. J. Acoust. Soc. Am. **98**(6), 3302–3308 (1995). https://doi.org/10.1121/1.413817
11. S. Bilbao, B. Hamilton, J. Botts, L. Savioja, Finite volume time domain room acoustics simulation under general impedance boundary conditions. IEEE/ACM Trans. Audio Speech Lang. Process. **24**(1), 161–173 (2016). https://doi.org/10.1109/TASLP.2015.2500018. Conference Name: IEEE/ACM Transactions on Audio, Speech, and Language Processing
12. T. Okuzono, T. Otsuru, R. Tomiku, N. Okamoto, A finite-element method using dispersion reduced spline elements for room acoustics simulation. Appl. Acoust. **79**, 1–8 (2014). https://doi.org/10.1016/j.apacoust.2013.12.010
13. J.A. Hargreaves, L.R. Rendell, Y.W. Lam, A framework for auralization of boundary element method simulations including source and receiver directivity. J. Acoust. Soc. Am. **145**(4), 2625–2637 (2019). https://doi.org/10.1121/1.5096171

14. F. Pind, A.P. Engsig-Karup, C.H. Jeong, J.S. Hesthaven, M.S. Mejling, J. Strømann-Andersen, Time domain room acoustic simulations using the spectral element method. J. Acoust. Soc. Am. **145**(6), 3299–3310 (2019). https://doi.org/10.1121/1.5109396
15. E.A. Lehmann, A.M. Johansson, Diffuse reverberation model for efficient image-source simulation of room impulse responses. IEEE/ACM Trans. Audio Speech Lang. Process. **18**(6), 1429–1439 (2009). https://doi.org/10.1109/TASL.2009.2035038
16. U. Kristiansen, A. Krokstad, T. Follestad, Extending the image method to higher-order reflections. Appl. Acoust. **38**(2–4), 195–206 (1993). https://doi.org/10.1016/0003-682X(93)90051-7
17. Z. Meng, K. Sakagami, M. Morimoto, G. Bi, K.C. Alex, Extending the sound impulse response of room using extrapolation. IEEE Trans. Speech Audio Process. **10**(3), 167–172 (2002). https://doi.org/10.1109/TSA.2002.1001981
18. G. Götz, S.J. Schlecht, V. Pulkki, in *Proc. Int. Conf. Immersive and 3D Audio: from Architecture to Automotive (I3DA)*. A dataset of higher-order Ambisonic room impulse responses and 3D models measured in a room with varying furniture (Online conference, 2021). https://doi.org/10.1109/i3da48870.2021.9610933
19. S. Kim, J.H. Yoo, J.W. Choi, Echo-aware room impulse response generation. J. Acoust. Soc. Am. **156**(1), 623–637 (2024)
20. S. Lee, H.S. Choi, K. Lee, in *Proc. IEEE Workshop Appl. Signal Process. Audio Acoust. (WASPAA)*. Yet another generative model for room impulse response estimation (New Paltz, 2023). https://doi.org/10.1109/WASPAA58266.2023.10248189
21. A. Ratnarajah, Z. Tang, R. Aralikatti, D. Manocha, in *Proceedings of the 30th ACM International Conference on Multimedia*. Mesh2ir: neural acoustic impulse response generator for complex 3D scenes (2022), pp. 924–933. https://doi.org/10.1145/3503161.3548253
22. N. Borrel-Jensen, A.P. Engsig-Karup, C.H. Jeong, Physics-informed neural networks for one-dimensional sound field predictions with parameterized sources and impedance boundaries. JASA Express Lett. **1**(12, paper no. 122402) (2021). https://doi.org/10.1121/10.0009057
23. N. Borrel-Jensen, S. Goswami, A.P. Engsig-Karup, G.E. Karniadakis, C.H. Jeong, Sound propagation in realistic interactive 3D scenes with parameterized sources using deep neural operators. Proc. Natl. Acad. Sci. **121**(2), e2312159120 (2024). https://doi.org/10.1073/pnas.2312159120. Publisher: Proceedings of the National Academy of Sciences
24. X. Karakonstantis, E. Fernandez-Grande, in *Proc. 10th Conv. European Acoust. Assoc. (Forum Acusticum)*. Room impulse response reconstruction using physics-constrained neural networks (Turin, 2023). https://doi.org/10.61782/fa.2023.0804
25. A. Luo, Y. Du, M. Tarr, J. Tenenbaum, A. Torralba, C. Gan, in *Proc. 35th Conf. Neural Inf. Process. Syst. (NeurIPS)*. Learning neural acoustic fields (Curran Associates, Inc., 2022), New Orleans, pp. 3165–3177
26. A. Richard, P. Dodds, V.K. Ithapu, in *Proc. IEEE Int. Conf. Acoust. Speech Signal Process. (ICASSP)*. Deep impulse responses: estimating and parameterizing filters with deep networks (Singapore, 2022), pp. 3209–3213. https://doi.org/10.1109/ICASSP43922.2022.9746135
27. S. Majumder, C. Chen, Z. Al-Halah, K. Grauman, in *Proc. 35th Conf. Neural Inf. Process. Syst. (NeurIPS)*. Few-shot audio-visual learning of environment acoustics (Curran Associates, Inc., 2022), New Orleans, pp. 2522–2536
28. K. Su, M. Chen, E. Shlizerman, in *Proc. 35th Conf. Neural Inf. Process. Syst. (NeurIPS)*. Inras: implicit neural representation for audio scenes (Curran Associates, Inc., 2022), New Orleans, pp. 8144–8158
29. A. Sarroff, R. Michaels, in *Proc. 23rd Int. Conf. Digital Audio Effects (DAFx)*. Blind arbitrary reverb matching (Online Conference, 2020), pp. 24–30
30. N. Singh, J. Mentch, J. Ng, M. Beveridge, I. Drori, in *Proc. IEEE/CVF Int. Conf. Computer Vision (ICCV)*. Image2reverb: cross-modal reverb impulse response synthesis (Montreal, 2021), pp. 286–295. https://doi.org/10.1109/ICCV48922.2021.00035
31. H. Kon, H. Koike, Estimation of late reverberation characteristics from a single two-dimensional environmental image using convolutional neural networks. J. Audio Eng. Soc. **67**(7/8), 540–548 (2019). https://doi.org/10.17743/jaes.2018.0069
32. S. Liang, C. Huang, Y. Tian, A. Kumar, C. Xu, AV-NeRF: Learning neural fields for real-world audio-visual scene synthesis. arXiv preprint arXiv:2302.02088 (2023). https://doi.org/10.48550/arXiv.2302.02088
33. J. Koo, S. Paik, K. Lee, Reverb conversion of mixed vocal tracks using an end-to-end convolutional deep neural network. arXiv preprint arXiv:2103.02147 (2021). https://doi.org/10.48550/arXiv.2103.02147
34. C.J. Steinmetz, V.K. Ithapu, P. Calamia, in *Proc. IEEE Workshop Appl. Signal Process. Audio Acoust. (WASPAA)*. Filtered noise shaping for time domain room impulse response estimation from reverberant speech (Online conference, 2021), pp. 221–225. https://doi.org/10.1109/WASPAA52581.2021.9632680
35. A.H. Moore, M. Brookes, P.A. Naylor, in *Proc. 21st European Signal Process. Conf. (EUSIPCO 2013)*. Room geometry estimation from a single channel acoustic impulse response (Marrakech, 2013) IEEE, NJ, USA
36. D. Markovic, F. Antonacci, A. Sarti, S. Tubaro, in *Proc. IEEE Workshop Appl. Signal Process. Audio Acoust. (WASPAA)*, Estimation of room dimensions from a single impulse response (New Paltz, 2013). https://doi.org/10.1109/WASPAA.2013.6701867
37. M. Kuster, Reliability of estimating the room volume from a single room impulse response. J. Audio Eng. Soc. **124**(2), 982–993 (2008). https://doi.org/10.1121/1.2940585
38. W. Yu, W.B. Kleijn, Room acoustical parameter estimation from room impulse responses using deep neural networks. IEEE/ACM Trans. Audio Speech Lang. Process. **29**, 436–447 (2020). https://doi.org/10.1109/TASLP.2020.3043115
39. J.A. Moorer, About this reverberation business. Comput. Music J. **3**(2), 13–28 (1979). https://doi.org/10.2307/3680280
40. R. Yamamoto, E. Song, J.M. Kim, in *Proc. IEEE Int. Conf. Acoust. Speech Signal Process. (ICASSP)*. Parallel WaveGAN: a fast waveform generation model based on generative adversarial networks with multi-resolution spectrogram (Online Conference, 2020), pp. 6199–6203. https://doi.org/10.1109/ICASSP40776.2020.9053795
41. K. Prawda, S.J. Schlecht, V. Välimäki, Calibrating the Sabine and Eyring formulas. J. Acoust. Soc. Am. **152**(2), 1158–1169 (2022). https://doi.org/10.1121/10.0013575
42. F. Klein, S.V. Amengual Garí, in *Proc. IEEE Int. Conf. Acoust. Speech Signal Process. (ICASSP)*. The R3VIVAL Dataset: repository of room responses and 360 videos of a variable acoustics lab (Rhodes Island, Greece, 2023). https://doi.org/10.1109/ICASSP49357.2023.10097257
43. J. Traer, J.H. McDermott, Statistics of natural reverberation enable perceptual separation of sound and space. Proc. Natl. Acad. Sci. U.S.A. **113**(48), E7856–E7865 (2016). https://doi.org/10.1073/pnas.1612524113
44. Z. Tang, R. Aralikatti, A.J. Ratnarajah, D. Manocha, in *ACM SIGGRAPH 2022 Conference Proceedings*. Gwa: a large high-quality acoustic dataset for audio processing (Association for Computing Machinery, 2022), Vancouver, pp. 1–9
45. I. Szöke, M. Skácel, L. Mošner, J. Paliesek, J. Černockỳ, Building and evaluation of a real room impulse response dataset. IEEE J. Sel. Top. Signal Process. **13**(4), 863–876 (2019). https://doi.org/10.1109/JSTSP.2019.2917582
46. L. Wright, N. Demeure, Ranger21: a synergistic deep learning optimizer. arXiv preprint arXiv:2106.13731 (2021). https://doi.org/10.48550/arXiv.2106.13731
47. G. Götz, R. Falcón Pérez, S.J. Schlecht, V. Pulkki, Neural network for multi-exponential sound energy decay analysis. J. Acoust. Soc. Am. **152**(2), 942–953 (2022). https://doi.org/10.1121/10.0013416
48. M.R. Schroeder, New method of measuring reverberation time. J. Acoust. Soc. Am. **37**(3), 1187–1188 (1965). https://doi.org/10.1121/1.1909343
49. H.P. Seraphim, Untersuchungen über die Unterschiedsschwelle exponentiellen Abklingens von Rauschbandimpulsen. Acta Acustica United Acustica **8**(4), 280–284 (1958)
50. C. Hold, T. McKenzie, G. Götz, S. Schlecht, V. Pulkki, Resynthesis of spatial room impulse response tails with anisotropic multi-slope decays. J. Audio Eng. Soc. **70**(6), 526–538 (2022). https://doi.org/10.17743/jaes.2022.0017

## Publisher's Note

Springer Nature remains neutral with regard to jurisdictional claims in published maps and institutional affiliations.